\documentclass[preprint,5p,authoryear]{elsarticle}

\usepackage{amssymb}
\usepackage{amsmath}

\usepackage{url}
\usepackage{placeins}
\usepackage{ulem}
\usepackage{siunitx}
\usepackage{gensymb}
\usepackage{multirow}
\usepackage{tabularx}
\usepackage{ragged2e}
\usepackage{array}

\journal{Cryogenics}

\begin{document}

\begin{frontmatter}



\title{The Simons Observatory: A Novel Design for Heat Strap at Sub-Kelvin Temperatures
} 

\author[label1]{John Orlowski-Scherer}
\author[label2,label3,label6]{Oorie Desai}
\author[label2,label3]{Daniel Lee}
\author[label2,label3]{Nicholas Galitzki}
\author[label1]{Alex Manduca}
\author[label1]{Mark Devlin}
\author[label1]{Dominic Horne}
\author[label2,label3]{Henry Nachman}
\author[label4]{Yuhan Wang}
\author[label5]{Edward J. Wollack}

\affiliation[label1]{organization={Department of Physics and Astronomy, University of Pennsylvania},
            addressline={209 South 33rd St},
            city={Philadelphia},
            postcode={19146},
            state={Pennsylvania },
            country={United States}}

\affiliation[label2]{organization={Department of Physics, University of Texas at Austin},
            addressline={ 2515 Speedway $\#$5.208},
            city={Austin},
            postcode={78712},
            state={Texas},
            country={United States}}
            
\affiliation[label3]{organization={Weinberg Institute for Theoretical Physics, Texas Center for Cosmology and Astroparticle Physics},
            addressline={ 2515 Speedway $\#$5.208},
            city={Austin},
            postcode={78712},
            state={Texas},
            country={United States}}

\affiliation[label4]{organization={Department of Physics, Cornell University},
            addressline={142 Sciences Dr},
            city={Ithaca},
            postcode={14853},
            state={New York },
            country={United States}}

\affiliation[label5]{organization={NASA Goddard Space Flight Center},
            addressline={8800 Greenbelt Road},
            city={Greenbelt},
            postcode={20771},
            state={Maryland},
            country={United States}}

\affiliation[label6]{organization={University of Florida},
            addressline={2001 Museum Road},
            city={Gainesville},
            postcode={32611},
            state={Florida},
            country={United States}}



\begin{abstract}

We present a novel process for the fabrication of copper heat straps and a measurement of their thermal performance at sub-Kelvin temperatures. The conductance of the heat straps is consistent with the bulk conductivity of annealed copper over the range 60\,mK to 1\,K. We find no evidence for deviation from a linear dependence of conductivity with temperature over the explored range, indicating that the contact resistance at the strap ends is sub-dominant. Heat straps developed and validated through the described process have been deployed in both the Large Aperture Telescope Receiver and a Small Aperture Telescope of the Simons Observatory and have met thermal requirements. 
\end{abstract}


\end{frontmatter}



\section{Introduction}
\label{sec:intro}
Cryogenic cooling has come to play a critical role in many scientific and industrial fields, such as superconducting systems \citep{Doi2002}, aerospace applications \citep{Homan2014}, direct dark mater searches~\citep{cdms2022}, and astronomy \citep{Galitzki2024, Zhu2021}. Cooling to temperatures below $100$\,mK generally requires either dilution refrigerators (DRs, $\lesssim100$\,mK) or sorption fridges \citep{Bhatia2000}. At these temperatures creating an effective thermal link between the cryocooler and the instrument space is challenging. The conductivity of all materials drops significantly below $4$\,K \citep{Woodcraft2009, Nachman2026}, and the contact resistance at material interfaces can become comparable to the material resistance \citep{Didschuns2004, Dhuley2019}. Moreover, mechanical compliance between the cryocooler and the instrument must be maintained, which is further complicated by potential differential thermal contraction between the two surfaces. Finally, sensitive measurement apparatuses need to be vibrationally isolated from potentially disruptive cryocoolers, such as pulse tubes (PTs). These problems have traditionally been solved by flexible heat straps manufactured from $99.99\%$ purity oxygen-free high conductivity (OFHC) copper. These heat straps consist of two blocks linked by braid, with the braid connected to the blocks by inserting the braid into a cavity in the block and then pressing the block.  

The Simons Observatory \citep[SO;][]{SOgoals} is a next generation cosmic microwave background (CMB) experiment which recently began observations in the Atacama desert \citep{Crowley2024}. SO comprises one Large Aperture Telescope \citep[LAT;][]{Zhu2021} and three Small Aperture Telescopes \citep[SATs;][designated SATp1, SATp2, and SATp3]{Galitzki2024}, each of which uses transition edge sensors \citep[TESes;][]{Stevens2020, McCarrick2021} to detect incoming photons. TESes in general operate in the $100$\,s of mK range, and the TESes used in SO have been designed to operate at or below $100$\,mK to minimize noise. All four telescopes use DRs\footnote{Bluefors, \url{https://bluefors.com/}} to achieve a base temperature of $80\,$mK at the DR mixing chamber (MXC). In order to achieve our goal TES temperature of $100$\,mK, the total thermal gradient from DR MXC to a TES is required to be $<20\,$mK over distances of up to $2\,$m under a maximum expected head load of $10\,\mu$W. As an initial solution, all four telescopes used commercially available OFHC heat straps manufactured by Technology Applications, Inc. (TAI).\footnote{Technology Applications, Inc. \url{ https://www.techapps.com/}}

The cryogenic performance of the TAI heat straps was measured between $0.13$ and $10$\,K in~\cite{Dhuley2017} (henceforth D17). D17 found that  the pressed boundary resistance between the strap ends and the strap braid began to dominate the strap conductance around $\sim 1.5$\,K.\footnote{Throughout this paper we use boundary resistance to mean this effect, and not the interface resistance between the strap and whatever it is being bolted to.} The boundary has a roughly quadratic relationship between conductance and temperature, which results in a rapid degradation in strap performance below $1$\,K. In particular D17 found that conductivity improvements of $\sim 5-10\times$ were possible in the $100$\,mK temperature range that the SO instruments operate at. This motivated us to design improved heatstraps, hereafter refered to as the SO heat straps, which are currently installed in SATp1 (as of February 2025) and the LAT (as of January of 2026). 

There are two routes to improving the performance of the heat straps: improving the bulk conductivity of the copper and reducing the thermal boundary resistance at the interface between the block and braid of the strap. At low temperatures the bulk thermal conductivity scales linearly with the temperature, while the conduction through boundaries scales as $T^{\beta}$ for $2\lesssim \beta \lesssim3$ typical~\citep{Gmelin1999}. Therefore, at very cold temperatures, the boundary resistance will dominate over the bulk conductivity, while at higher temperatures the bulk resistance will be the primary contribution. The turn-over temperature between the regimes is a product of the quality of the boundary; the better the boundary contact, the lower the turn-over temperature. The first goal of strap design is therefore to make a boundary of such quality that the turn-over temperature is below the temperature range of use for the heat strap. The second goal is to increase the bulk conductivity as much as possible, in order to maximize the conductance of the strap, which can be done by annealing the copper to reduce the number of dislocations in the crystaline structure, which are sites for phonon scattering~\citep[see Section~\ref{sec:disc} and ][for an extensive review]{simon1992}.

The SO heat straps have significantly higher thermal conductivity than the TAI strap, with the SO strap conductivity being consistent with the higher range of literature values for vacuum annealed OFHC copper. Furthermore, from lab measurements, we find the conductance of the SO heat straps to be linear with temperature down to $60$\,mK, indicating that the bulk conductivity of copper dominates over boundary resistance over the entire temperature range.


This paper is structured as follows. In Section~\ref{sec:desg}, we describe the design and manufacturing process for these heat straps. In Section~\ref{sec:thermal}, we describe the in-lab and in-field performance of the SO straps. In Section~\ref{sec:disc}, we interpret the results of out thermal testing and compare their performance to that of the TAI straps. Finally, in Section~\ref{sec:conc}, we give some conclusions and suggestions for future improvements.

\section{Heat Strap Design and Manufacture}
\label{sec:desg}

\begin{figure}
\centering
\includegraphics[width=0.85\columnwidth]{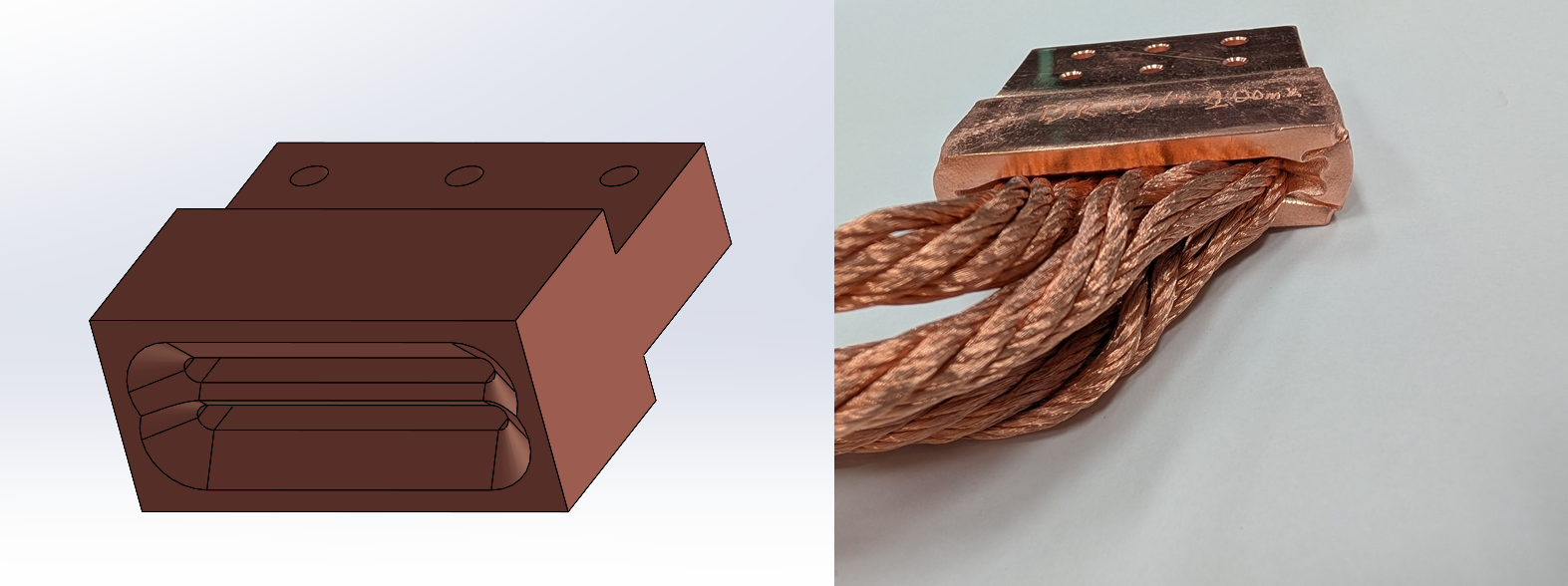}
\caption{The solid model for a strap block (left) and real pressed strap (right). The block is initially machined to match the model on the left, with a step in it. That step is then pressed to 50 tons and machined flat.}
\label{fig:block}
\end{figure}

Each SO heat strap comprises two blocks connected by braid. The blocks are solid machined OFHC copper blocks connected by OFHC braid (Fig~\ref{fig:block}).\footnote{Throughout, we used ``block'' to refer specifically to this part before pressing, and ``end'' to refer to the combined strap/block after pressing.} The amount of braid sets the effective $\frac{\text{area}}{\text{length}}$ (A/L), which in turn determines the absolute thermal conductivity. Each block has a machined slot into which the braid is inserted. This assembly is compressed under high pressure and the two materials are joined. The machined slot is chamfered and filleted to avoid stress concentrations where the braid leaves the slot. The region where the braid is inserted is stepped up in thickness so that only the raised step is deformed during the pressing process, leaving the bolted thermal interface roughly flat (See Figure~\ref{fig:block}). The blocks are then machined completely flat and mounting holes are machined in them. During prototyping we found that the positions of the holes shifted during pressing, so machining the holes after pressing ensures the correct pattern. The braid is OFHC copper, $1/2$\,inch in diameter, comprised of 38 gauge wire. The slot width is standardized to $44$\,mm, which fits two braids of 7 strands when fanned out. The slot width can be proportionately adjusted to accept more braid. 

The manufacturing process begins by machining the blocks, which was done on a 5-axis computer numerical control (CNC) machine, and cutting the braid to length\footnote{The braid was cut somewhat long and then trimmed to length immediately before pressing.}. The blocks were cleaned after machining using a mixture of Simple Green\footnote{\url{https://simplegreen.com/products/all-purpose-cleaner/}} and water to remove grease. Both the blocks and the braid were then annealed before pressing; 
we call this step pre-press annealing. 
The purpose of the pre-press annealing is to soften the copper, so that both the block and the strap deform as much as possible during the pressing process. The parts were pre-annealed under vacuum at a pressure of $\lesssim 5$\,mbar for four hours at $450\degree$\,C and allowed to cool slowly over 18 hours. The $450\degree$\,C temperature was chosen based off of a literature survey of common annealing temperatures~\citep{Karasev2012}. All annealing was performed in-house using an Across International 1200C Furnace.\footnote{\url{https://www.acrossinternational.com/1200c-controlled-atmosphere-muffle-furnace-etl-ce-certified.html}}

During prototyping, some parts developed oxidization while annealing. It was discovered that further annealing at $650\degree$\,C removed the oxidization layer. As such, one prototype, called Prototype B, was made, which was annealed at $650\degree$\,C. It was found to have better conductivity than the unannealed straps but worse conductivity than those annealed at $450\degree$\,C (compare \cite{Karasev2012}). As such, we recommend annealing at $450\degree$\,C, but annealing at $650\degree$\,C can recover oxidized parts and still provide acceptable conductivity. See Section~\ref{sec:disc} for a full discussion of the dependence of thermal conductivity with annealing temperature. 

After pre-press annealing, the braid was inserted into the block and the block was then crushed in a hydraulic press to 50 tons of force\footnote{The limit of our press}; this is the pressing step. The pressing surface area is $18.75\,\text{cm}^2$, resulting in a pressure of $\sim240$\,MPa. The yield strength for fully annealed OFHC is $\sim70$\,MPa while for full hardened OFHC it is $\sim305$\,MPa \citep{ASM-copper}, indicating the necessity of the pre-annealing step. The pressing was then repeated for the other end of the strap. After pressing, the ends of the strap were machined flat and the through holes drilled. Finally, the strap was re-annealed using the same process detailed above; we call this the post-annealing. The post-annealing temperature matched that of the pre-press annealing. The copper has significantly work hardened after pressing and machining, and the post-annealing restores the high thermal conductivity of the copper. Work hardening is the process whereby dislocations accumulate in a metal that is plastically deformed \citep{Silberschmidt2024}. These dislocation create sites for phonon scattering, which increase the thermal conductivity of the material, and they can be ameliorated by annealing \citep{Humphreys2024}. After manufacture, effort was made to minimize work hardening of the straps: they were packed in foam for storage and shipping.


\begin{figure}[h!]
\centering
\includegraphics[width=0.85\columnwidth]{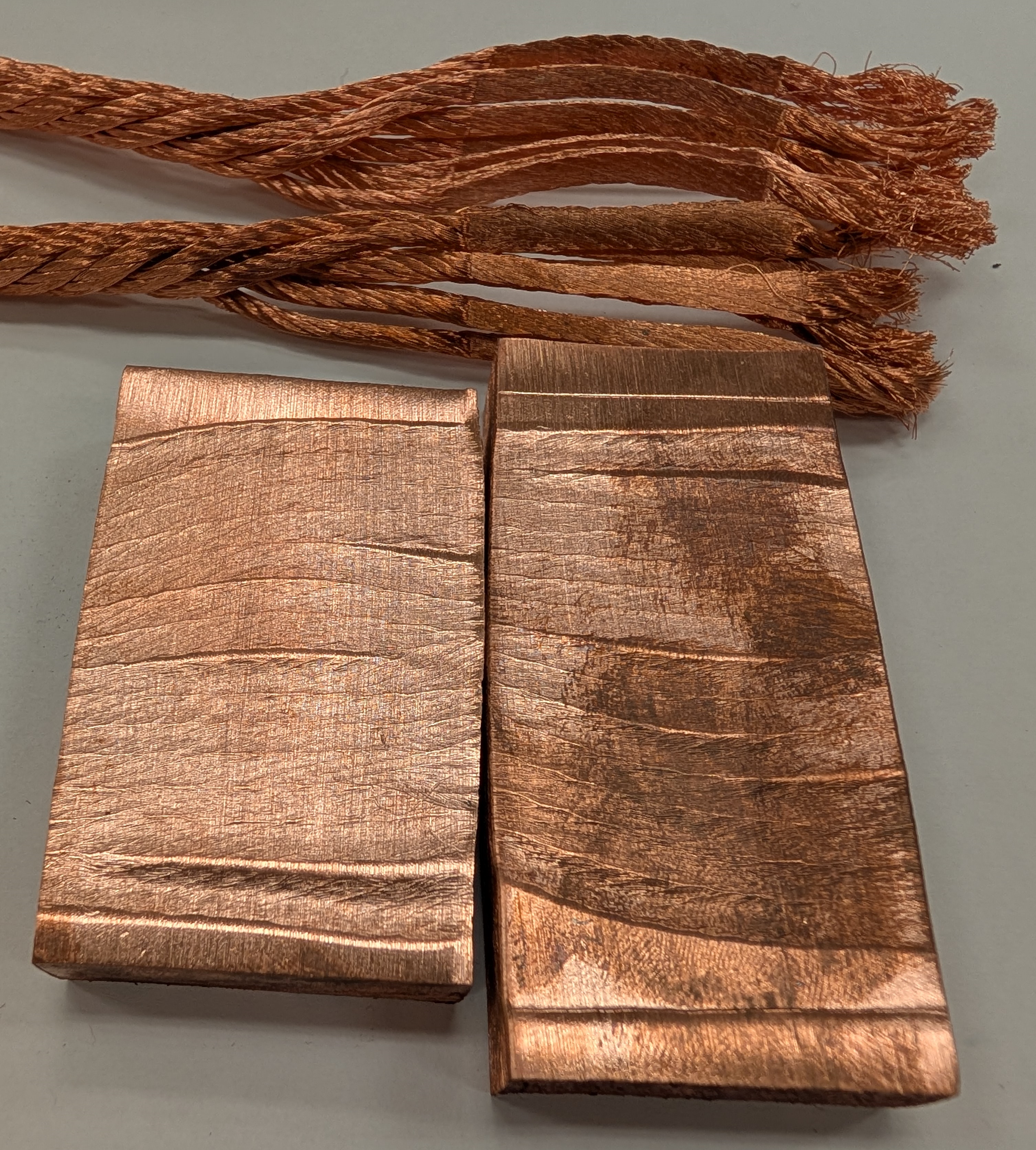}
\caption{
A pair of separate blocks and strap which have been pre-annealed and pressed as in Section~\ref{sec:desg}. Indentations from individual copper wires can clearly be seen in the copper block. The contact between the block and the braids is exceptional, with no gaps, although they have not fused into one part. The oxidation seen on the right block is from sitting exposed for several months and not indicative of the condition of the piece when pressed.}
\label{fig:cut_end}
\end{figure}

To visually inspect the quality of the pressed interface, we cut open a prototype strap end. This test piece went through the full manufacturing process described above, including pre-press annealing, pressing, and post annealing. Visually, the cut piece reveals the copper has plastically deformed to create gap-free contact between the strap and block; see the top of Figure~\ref{fig:cut_end}. The boundary between the two parts is, however, still visible, and closer inspection reveals that the part is indeed only pressed together and has not cold-welded. As a further test, we pressed a strap between two separate copper blocks; the resulting assembly could be pulled apart to inspect the strap-block interface. Both the blocks and strap were pre-annealed and pressed but not post-annealed. The resulting interface shows the clear imprint of individual wires from the braid on the block pieces; see the bottom of Figure~\ref{fig:cut_end}, indicating excellent pressed contact but reinforcing that the parts have not actually cold welded or in any way fused.


\section{Thermal Testing}
\label{sec:thermal}
To evaluate whether the manufactured heat straps meet our thermal performance objectives, we performed cryogenic validation at temperatures between 60\,mK and 1\,K. The results directly informed the final design implemented in the Simons Observatory cryogenic receivers.

\subsection{Methodology}
\begin{figure}
\centering
\includegraphics[width=\columnwidth]{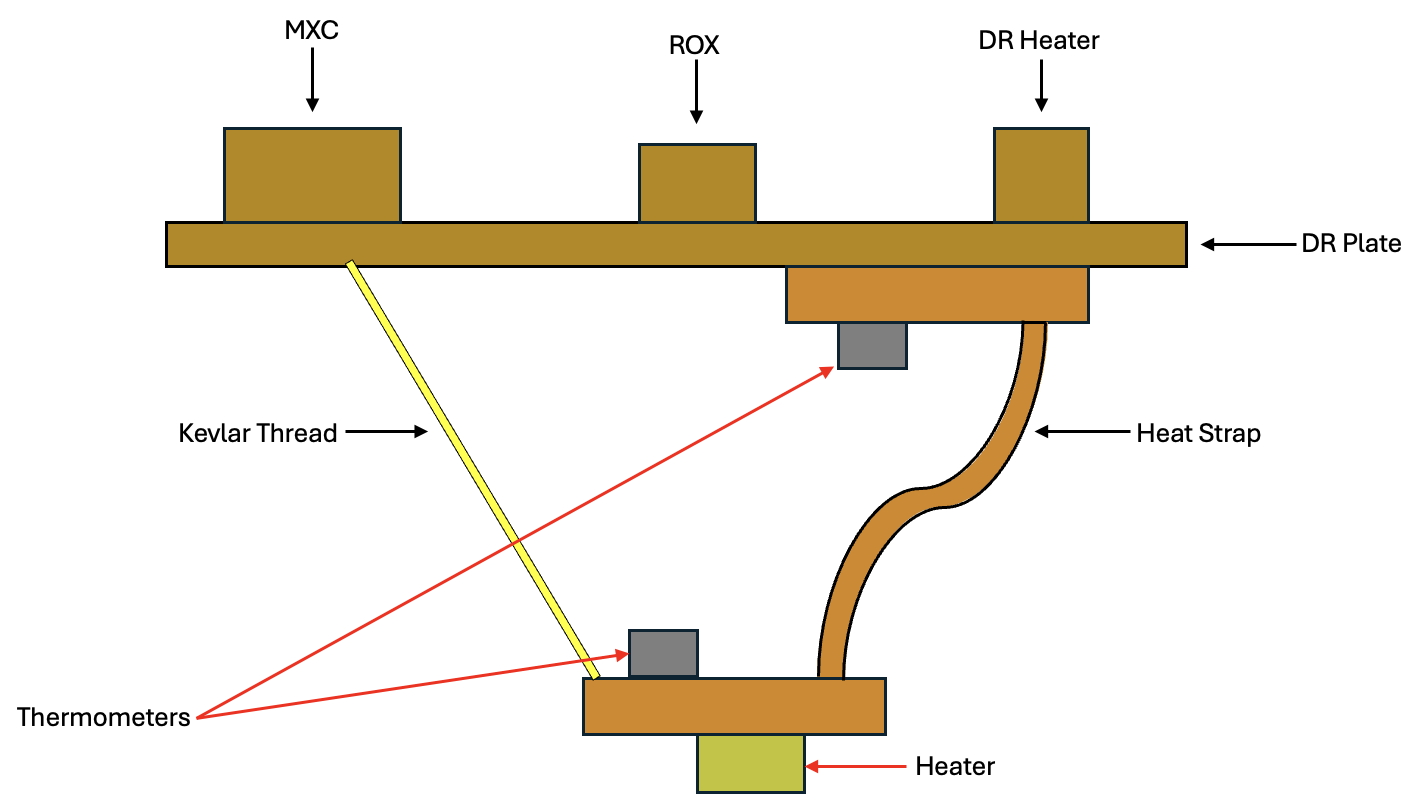}
\caption{This figure is a schematic of the heat strap attached to the MXC of the DR and shows the placements of the thermometers and the heater on a heat strap. The yellow rectangle underneath the strap is the heater and the two gray squares are the thermometers. The Kevlar thread supports one end of the heat strap while contributing minimal parasitic thermal loading. The DR heater is used to PID the one end of the strap to a constant temperature while the other heater applies a thermal load to the other end.}
\label{fig:dr_setup}
\end{figure}

We follow the method described in \cite{Nachman2026} to measure the thermal conductivity, 
 with a thermometer on each end of the sample and a heater to drive the gradient. We attach one end of the heat strap to the MXC flange of a Bluefors DR with a 100\,$\Omega$ resistive heater attached to the other end of the strap (see Figure~\ref{fig:dr_setup}). 
Voltage setpoints are output from a  Keithley 2280S power supply which also provides a measurement of the current going through the heater. The temperatures are obtained with two Lake Shore calibrated Ruthenium Oxide (ROX) sensors. The thermometers are read out with the Bluefors Temperature Controller (TC) which provides AC resistance bridge measurements with an associated scanner that cycles between temperature channels. During testing, we actively monitored the two ROXs attached to the sample as well as the MXC thermometer with 5\,s wait times after channel switches and 12\,s sample periods. The TC also provided the Proportional-Integral-Derivative (PID) control of the MXC. The Observatory Control System \cite[OCS;][]{Koopman2020}  software provided device control and data acquisition capabilities. 

\begin{table*}[th]
\centering
\begin{tabularx}{\textwidth}{|>{\RaggedRight\arraybackslash}p{3.3cm}|
                              >{\Centering\arraybackslash}p{1.2cm}|
                              >{\Centering\arraybackslash}X|
                              >{\Centering\arraybackslash}p{1.5cm}|
                              >{\RaggedRight\arraybackslash}p{2.5cm}|
                              >{\RaggedRight\arraybackslash}X|
                              >{\RaggedRight\arraybackslash}p{2.0cm}|}
\hline
\textbf{Heat Strap} & \textbf{Length (mm)} & \textbf{Single Strand Cross-section 
($mm^{2}$)} & \textbf{Number of Strands} & \textbf{Preparation} & \textbf{Power Law Fit ($\beta$)} & \textbf{Estimated RRR}  \\
\hline
TAI & 150 & $3.2 $ & 36 & Unmodified & $1.18 \pm 0.01$ & $45.2 \pm 1.5$* \\
\hline
TAI Welded and Annealed & 150 & $3.2$ & 36 & E-beam welded,  $450\degree$\,C annealed & $1.03 \pm 0.01$ & $197.3 \pm 1.2$ \\
\hline
Prototype A (Test 1) & 200 & $3.1 $ & 45 & $450\degree$\,C annealed & $1.06 \pm 0.01$ & $222.2 \pm 1.9$\\
\hline
Prototype A (Test 2) & 200 & $3.1$ & 9 &  $450\degree$\,C annealed & $0.97 \pm 0.02$ & $227.4 \pm 2.6$\\
\hline
Prototype B & 708 & $3.1$ & 14 &  $650\degree$\,C annealed & $1.00 \pm 0.01$ & $158.9 \pm 0.2$\\
\hline
Prototype C & 229 & $3.8$ & 12 & Unannealed& $1.02 \pm 0.01$ & $45.7 \pm 0.1$\\
\hline
SO SAT Strap 1 & 280 & $3.1$ & 21 &  $450\degree$\,C annealed & $1.02 \pm 0.02$ & $235.7 \pm 1.9$\\
\hline
SO SAT Strap 2 & 280 & $3.1$ & 21 &  $450\degree$\,C annealed & $0.99 \pm 0.02$ & $232.9 \pm 1.3$\\
\hline
\end{tabularx}
\caption{
Strap specifications and the resultant fit parameters for the exponential fit and for the effective RRR value. 
The Prototype A strap was tested twice, the second time with all but 9 of the copper braids cut to allow us to extend the testing range to lower temperatures while also providing a consistency check of our ability to recover similar performance values over two distinct cool downs. We also validated both straps for the deployed SO SAT system which are shown here and demonstrate the repeatability of the manufacturing process. *The RRR estimate assumes $\beta=1$, resulting in a less reliable result for values that deviate significantly.}
\label{table:1}
\end{table*}

To build off the work of \cite{Dhuley2017}, we tested commercial straps from TAI provided by TAI. We tested TAI straps both unmodified, and after electron-beam (e-beam) welding and annealing (first two rows of Table~\ref{table:1}). E-beam welding uses a concentrated beam of high-energy electrons to fusion weld metal parts. Critically for us, e-beam welds can be ``full-depth'', meaning they pass entirely through the parts being joined, resulting in excellent thermal conductivity between the parts. The annealing was performed as discussed in Section~\ref{sec:desg}, although only the post-annealing was done as the straps were delivered to us from TAI already pressed. 

We also tested a number of SO heat straps, Prototype A, B, and C as well as SO SAT Strap 1 and 2 in Table~\ref{table:1}. Each of the SO heat straps was made with the same braid, although in differing number of braids, and with the same block geometry, except that the length of the slot in the block was changed to accommodate the differing number of strands. The bolting patterns and bolted surface area also differed but do not affect these measurements as these surfaces are not involved in pressing and the thermometers were mounted directly to the braid ends, so the bolt pattern used to attach the braid to the DR made no difference. Between Test 1 and Test 2, a number of the strands were cut as a check on the effect of changing the A/L. Prototype B had fewer strands than A but more importantly was annealed at $650\degree$\,C as opposed to A, which was annealed at $450\degree$\,C. Prototype C was pressed to $50$\,T but not annealed. Prototype C also used a strand with a different cross-section from the other prototypes and SAT straps. SO SAT Straps 1 and 2 are the actual straps that were deployed to SATp1. They mirror Prototype A in construction and annealing, only differing in the exact number of strands, and were tested as their deployment schedule permitted it.

\subsection{Results}
\begin{figure*}[htbp]
    \centering
    \includegraphics[width=\textwidth]{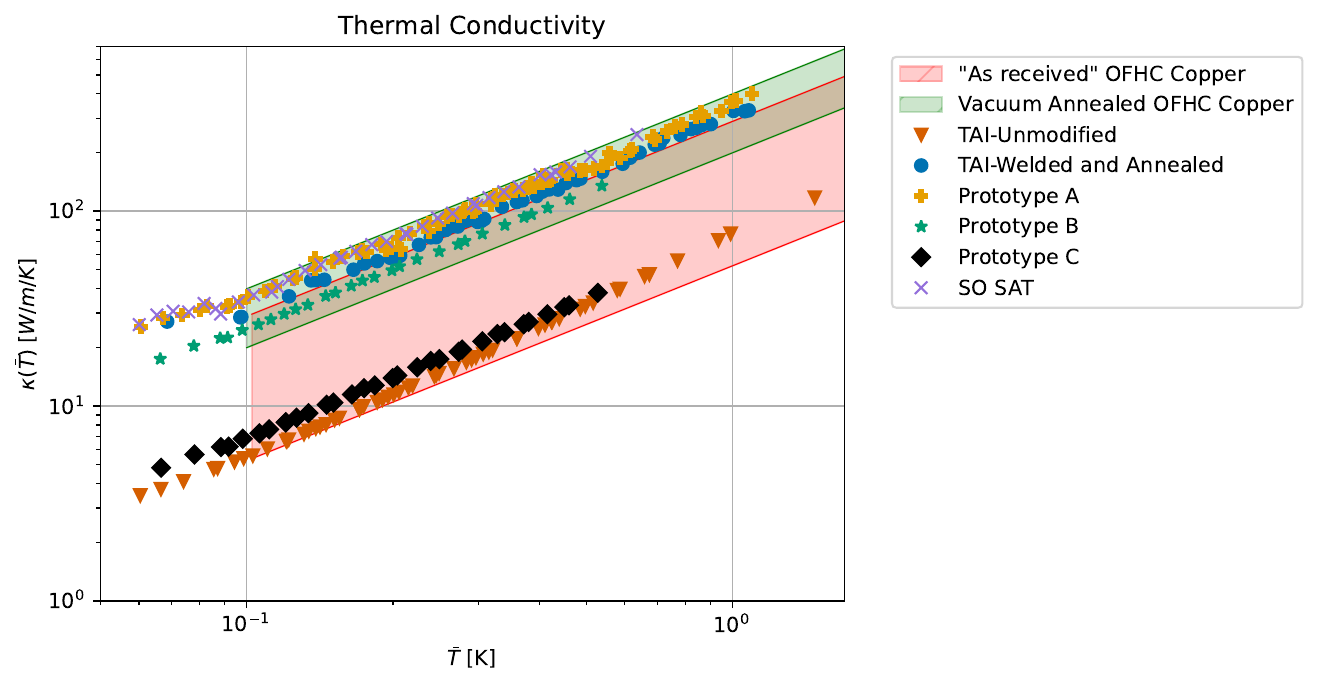}
    \caption{Thermal conductivity data of the tested heat straps with typical ranges for ``as received'' OFHC copper (red) and vacuum annealed OFHC copper (green) \citep{Woodcraft2009}. Prototype A includes the combined data from both tests. Only SO SAT Strap 1 data is shown.}
    \label{fig:conductivity}
\end{figure*}

We obtained the thermal conductivity of the straps at low temperatures and fit the resultant data to a power law,
\begin{equation}
    k = \alpha T^\beta.
\end{equation}
A $\beta$ value of $\sim1$ indicates the heat strap is behaving like pure copper in this regime versus higher powers that could be indicative of a contact resistance in the strap impacting the conductivity. For straps that have a value close to 1, we extract an effective residual-resistance ratio (RRR) value by fitting the data using the copper parameters found in  \cite{hust_lankford_1984}. 
The resultant RRR value quantizes the relative conductivity of a copper sample, with higher values indicative of increased thermal conductivity. The derived RRR values are shown in Table \ref{table:1} along with associated one sigma uncertainties. 
 Figure \ref{fig:conductivity} shows the plot of thermal conductivity vs. temperature for the straps that were tested.

\section{Discussion}
\label{sec:disc}
When analyzing the results of the thermal conductivity measurements made in Section~\ref{sec:thermal}, there are two variables at play: the annealing process and the joining process (i.e. pressing or e-beam welding). We expect the annealing process to improve the bulk conductivity of the copper, while the joining process should dictate the boundary resistance between the block and braid. In annealing, metal is heated above its recrystalization temperature, reducing the number of dislocations and therefore increasing its thermal conductivity \citep{rauta2016}. Annealing thus improves the conductivity in the linear, high temperature regime \citep[$\gtrsim 1.6$\,K][]{Dhuley2017}. The different joining methods affect the boundary between the blocks and the braid, and hence the boundary thermal resistance. The boundary thermal resistance results in thermal conductivity that is quadratic in temperature, and dominates at lower temperatures \citep[$\lesssim 1.5$\,K][]{Dhuley2017}. The exact turn-over temperature between bulk and boundary resistance depends on the details of the annealing and joining process.

\subsection{Effect of Joining Method}
\label{sec:disc:ebeam}
For all strap samples, the dependence of the conductance with temperature is linear, and hence the dominant contribution to thermal resistance is the bulk resistance of the copper. We see no evidence of quadratic behavior that would be associated with dominance by the boundary resistance of the braid/block interface. In the case of the unmodified TAI strap, this stands in contrast with \cite{Dhuley2017}, which found a quadratic dependence on temperature below 1.5\,K. The source of this difference is unclear. It may be that TAI changed their manufacturing process between the Dhuley paper ($\sim2017$) and when our strap was manufactured ($\sim 2019$). Alternatively, the TAI manufacturing process may produce inconsistent results at low temperature. In any case, we find all of the joining methods (TAI, SO pressing, or e-beam welding) result in boundary thermal resistance which is sub-dominant to the bulk resistance. This is not to say that there are no differences in boundary thermal resistance between the joining processes, only that any difference in boundary resistance is too small to be measured. Future research into the differences in boundary resistance between these methods is warranted, however for the purposes of manufacturing heat straps as described in this paper, all methods are viable. 

Additionally, the joining method has no effect on the RRR of the straps. The RRR of the welded and annealed sample (TAI Welded and Annealed, 197) is somewhat but not significantly lower than the unwelded annealed samples (RRR$~\sim 220$). As such, we find no evidence that e-beam welding improves the thermal performance of the straps, although given that we only have one comparative data point (the Unmodified TAI strap) neither can we reject that hypothesis. The e-beam welding having no effect is consistent with not seeing evidence of boundary resistance in any of the straps.

\subsection{Effect of Annealing}
\label{sec:disc:ann}
The RRR of the sample straps was substantially improved by annealing, while again there was no impact on the slope of the power law fit. The samples annealed at $450\degree$\,C all had RRRs in the range of 200-230, comparable to solid vacuum annealed copper~\citep{hust_lankford_1984} and a factor of $\sim 5$ higher than the unannealed samples (RRR$\sim45$). This is the highest performance possible without using higher purity copper. The one sample annealed at $650\degree$\,C, Prototype B, has a somewhat lower RRR, $\sim160$, than those annealed at $450\degree$\,C, but still significantly higher than the unannealed samples. The unannealed straps have RRR's which are consistent with the mid to low end of vacuum annealed samples and the high end of  ``as received'' \citep{Valois2024}. 

The details of the annealing process affect the resulting RRR. While we only had one sample (Prototype B) annealed at $650\degree$\,C, it had a statistically significantly lower RRR. Qualitatively, the braids of this sample had a notably ``crunchy'' feel post annealing, as did other untested samples which were annealed at that temperature. Literature on the impact of annealing temperature on RRR is limited: \cite{Karasev2012} found that copper wire samples peaked in RRR around $450\degree$\,C and declined $\sim 20\%$ at $650\degree$\,C, inline with the findings of this paper. They did, however, also find significant variance between samples both in the peak temperature and variation between $450$ and $650\degree$\,C. Moreover, \cite{Dong2023} found the work hardening rate of copper to increase when the copper was annealed beyond $\sim400\degree$\,C. They proposed increased growth of annealing twins as the mechanism for increased work hardening. Annealing twins are pairs of crystalline grains which differ only by a reflection across some common plane. They are formed during recrystallization after annealing~\citep[p.\,897,][]{PhysicalMetallurgy}, and they could negatively impact the thermal conduction at low temperatures, as they introduce boundaries for phonon scattering. Alternatively, Prototype B could simply have work hardened more than the samples annealed at a lower temperature.  While every effort was made to minimize work hardening of our samples during transport and testing, it is possible that work hardening reduced the RRR of Prototype B. Future research systematically investigating the impact of annealing process details on thermal conductivity would be productive.

\subsection{In-Field Performance}
The two validated SO SAT heat straps were deployed on one of the three SATs in February of 2025 to improve the thermal performance of the instrument. Prior to the installation of the straps, there was a 37\,mK gradient from the focal plane to the DR. The path included TAI flexible straps, an unannealed copper rod that provided a conductive path through several magnetic shielding layers, and associated mating surfaces. The combination of new SO Heat straps and a newly annealed rod provided a decrease in the gradient of 15\,mK, which is consistent with our predictions

Heat straps using the Prototype B parameters, pressed to 50 tons and annealed for four hours at $650\degree$\,C, were deployed to the SO LAT Receiver (LATR) in January of 2026 as part of the Advanced SO deployment \citep[ASO;][]{OrlowskiScherer27}. We were able to measure the gradients across these straps in-situ; the average gradient from the MXC to the TESes is $5.5$\,mK, with a simple standard deviation of $3.8$\,mK and a maximum gradient of $15.0$\,mK. For comparison, with the old thermal bus solution installed in SO before the ASO upgrade, the typical gradient from the mixing chamber to the TESes was $20$\,mK. The new strapping solution therefore represents a factor $\sim4$ improvement in gradient for the average optics tube and is critical for achieving the ASO goals. Without this improvement, the TESes would operate significantly hotter than their goal temperature of $100$\,mK, incurring a noise penalty.

\section{Conclusion}
\label{sec:conc}

In this paper we have presented the design and testing of a novel heat strap, which provides superior performance to the state of the art at $\lesssim1.5$\,K. The new process differs from commercially available straps by the pressing technique used to combine the braid and strap ends and by the annealing of the strap. From the testing performed, it seems that annealing is more important for the low temperature conductivity than the joining procedure, although a direct comparison of annealed-but-not-pressed to pressed-but-not-annealed is needed to say with certainty. The pressed-and-annealed straps have a measured RRR a factor of $\sim5$ higher than representative commercial straps and consistent with bulk annealed copper~\citep{hust_lankford_1984}. Moreover, the power law fit to the conductivity for all straps is consistent with linear and shows no sign of a break in power law index. Notably this result differs from \cite{Dhuley2017}, who found that the TAI strap was dominated by the boundary resistance  below 1.5\,K. For all measured straps, every indication is that the conductivity is dominated by bulk conductivity of the copper, rather than the interface between the strap end and the braid.

Further improvements to the performance of these straps would require improving the conductivity of the copper itself. One simple method is to use higher purity copper. 5\,N copper is roughly an order of magnitude more conductive than the OFHC copper used in our straps. 5\,N copper is, however, also very soft and has a low tensile strength, which negatively impacts the durability of the strap. 
Variation in the annealing process could further increase the thermal conductivity of our straps by, for example, using oxygen annealing~\citep{Woodcraft2005} or fine tuning our current process. Indeed, we observe variations thermal conductivity between Prototype B and Prototype A; B was annealed at $650^\circ$\,C, while A was annealed at $450^\circ$\,C. This annealing difference resulted in a $\sim30\%$ reduction in the RRR of Prototype B as compared to Prototype A. 
Additional development of heat strap production methods and validation at cryogenic temperatures would be of great use to the field of cryogenic engineering. With upcoming projects in astronomy, superconducing, quantum computing,  and even axion searches pushing the capabilities of cryogenic systems, maximizing thermal performance at every turn will be more important than ever. 

\section{Acknowledgments}
This work was funded by grants from the Simons Foundation [MPS-Observatory-00457687, B.K.] and Simons Foundation International [SFI-MPS-Observatory-00007127, B.K.]. SO operates in the Parque Astronómico Atacama in northern Chile under the auspices of the Agencia Nacional de Investigación y Desarrollo (ANID). We thank the Republic of Chile for hosting SO in the northern Atacama, and the local indigenous Licanantay communities, whom we follow in observing and learning from the night sky. Some of the software developed to produce this analysis can be found publicly available on the Simons Observatory GitHub organization (https://github.com/simonsobs).

\bibliographystyle{elsarticle-harv} 
\bibliography{bibliography}

@INPROCEEDINGS{Valois2024,
       author = {{Valois}, J.~J. and {Nellis}, G.~F. and {Pfotenhauer}, J.~M.},
        title = "{Characterization of the thermal properties of OFHC copper at cryogenic temperature}",
    booktitle = {Materials Science and Engineering Conference Series},
         year = 2024,
       series = {Materials Science and Engineering Conference Series},
       volume = {1301},
        month = may,
    publisher = {IOP},
          eid = {012167},
        pages = {012167},
          doi = {10.1088/1757-899X/1301/1/012167},
       adsurl = {https://ui.adsabs.harvard.edu/abs/2024MS&E.1301a2167V}
}

@book{ASM-copper,
    author = {{Davis}, Joseph},
    title = {ASM Specialty Handbook - Copper and Copper Alloy},
    publisher = {ASM International},
    year = {2001}
}

@ARTICLE{Bhatia2000,
       author = {{Bhatia}, R.~S. and {Chase}, S.~T. and {Edgington}, S.~F. and {Glenn}, J. and {Jones}, W.~C. and {Lange}, A.~E. and {Maffei}, B. and {Mainzer}, A.~K. and {Mauskopf}, P.~D. and {Philhour}, B.~J. and {Rownd}, B.~K.},
        title = "{A three-stage helium sorption refrigerator for cooling of infrared detectors to 280 mK}",
      journal = {Cryogenics},
         year = 2000,
        month = jan,
       volume = {40},
       number = {11},
        pages = {685-691},
          doi = {10.1016/S0011-2275(00)00072-2},
       adsurl = {https://ui.adsabs.harvard.edu/abs/2000Cryo...40..685B}
}

@ARTICLE{cdms2022,
       author = {{SuperCDMS Collaboration} and {Albakry}, M.~F. and {Alkhatib}, I. and {Amaral}, D.~W.~P. and {Aralis}, T. and {Aramaki}, T. and {Arnquist}, I.~J. and {Ataee Langroudy}, I. and {Azadbakht}, E. and {Banik}, S. and {Bathurst}, C. and {Bauer}, D.~A. and {Bhattacharyya}, R. and {Brink}, P.~L. and {Bunker}, R. and {Cabrera}, B. and {Calkins}, R. and {Cameron}, R.~A. and {Cartaro}, C. and {Cerdeno}, D.~G. and {Chang}, Y. -Y. and {Chaudhuri}, M. and {Chen}, R. and {Chott}, N. and {Cooley}, J. and {Coombes}, H. and {Corbett}, J. and {Cushman}, P. and {De Brienne}, F. and {Dharani}, S. and {di Vacri}, M.~L. and {Diamond}, M.~D. and {Fascione}, E. and {Figueroa-Feliciano}, E. and {Fink}, C.~W. and {Fouts}, K. and {Fritts}, M. and {Gerbier}, G. and {Germond}, R. and {Ghaith}, M. and {Golwala}, S.~R. and {Hall}, J. and {Hassan}, N. and {Hines}, B.~A. and {Hollister}, M.~I. and {Hong}, Z. and {Hoppe}, E.~W. and {Hsu}, L. and {Huber}, M.~E. and {Iyer}, V. and {Jastram}, A. and {Kashyap}, V.~K.~S. and {Kelsey}, M.~H. and {Kubik}, A. and {Kurinsky}, N.~A. and {Lawrence}, R.~E. and {Lee}, M. and {Li}, A. and {Liu}, J. and {Liu}, Y. and {Loer}, B. and {Lukens}, P. and {MacFarlane}, D.~B. and {Mahapatra}, R. and {Mandic}, V. and {Mast}, N. and {Mayer}, A.~J. and {Theenhausen}, H. Meyer zu and {Michaud}, E. and {Michielin}, E. and {Mirabolfathi}, N. and {Mohanty}, B. and {Nagorny}, S. and {Nelson}, J. and {Neog}, H. and {Novati}, V. and {Orrell}, J.~L. and {Osborne}, M.~D. and {Oser}, S.~M. and {Page}, W.~A. and {Partridge}, R. and {Pedreros}, D.~S. and {Podviianiuk}, R. and {Ponce}, F. and {Poudel}, S. and {Pradeep}, A. and {Pyle}, M. and {Rau}, W. and {Reid}, E. and {Ren}, R. and {Reynolds}, T. and {Roberts}, A. and {Robinson}, A.~E. and {Saab}, T. and {Sadoulet}, B. and {Saikia}, I. and {Sander}, J. and {Sattari}, A. and {Schmidt}, B. and {Schnee}, R.~W. and {Scorza}, S. and {Serfass}, B. and {Poudel}, S.~S. and {Sincavage}, D.~J. and {Stanford}, C. and {Street}, J. and {Sun}, H. and {Thasrawala}, F.~K. and {Toback}, D. and {Underwood}, R. and {Verma}, S. and {Villano}, A.~N. and {von Krosigk}, B. and {Watkins}, S.~L. and {Wen}, O. and {Williams}, Z. and {Wilson}, M.~J. and {Winchell}, J. and {Wyko}, K. and {Yellin}, S. and {Young}, B.~A. and {Yu}, T.~C. and {Zatschler}, B. and {Zatschler}, S. and {Zaytsev}, A. and {Zhang}, E. and {Zheng}, L. and {Zuber}, S.},
        title = "{A Strategy for Low-Mass Dark Matter Searches with Cryogenic Detectors in the SuperCDMS SNOLAB Facility}",
      journal = {arXiv e-prints},
         year = 2022,
        month = mar,
          eid = {arXiv:2203.08463},
        pages = {arXiv:2203.08463},
          doi = {10.48550/arXiv.2203.08463},
archivePrefix = {arXiv},
       eprint = {2203.08463},
 primaryClass = {physics.ins-det},
       adsurl = {https://ui.adsabs.harvard.edu/abs/2022arXiv220308463S}
}

@INPROCEEDINGS{Crowley2024,
       author = {{Crowley}, Kevin T. and {Arnold}, Kam and {Galitzki}, Nicholas and {Gerras}, Remington G. and {Johnson}, Bradley R. and {Keating}, Brian and {Kusaka}, Akito and {Lee}, Adrian and {Nakata}, Hironobu and {Randall}, Michael J. and {Shaw}, Elle C. and {Terasaki}, Tomoki},
        title = "{The Simons Observatory: calibration and characterization of the first small-aperture telescope}",
    booktitle = {Millimeter, Submillimeter, and Far-Infrared Detectors and Instrumentation for Astronomy XII},
         year = 2024,
       editor = {{Zmuidzinas}, Jonas and {Gao}, Jian-Rong},
       series = {Society of Photo-Optical Instrumentation Engineers (SPIE) Conference Series},
       volume = {13102},
        month = aug,
          eid = {1310219},
        pages = {1310219},
          doi = {10.1117/12.3020564},
       adsurl = {https://ui.adsabs.harvard.edu/abs/2024SPIE13102E..19C}
}

@ARTICLE{Dhuley2017,
       author = {{Dhuley}, R.~C. and {Ruschman}, M. and {Link}, J.~T. and {Eyre}, J.},
        title = "{Thermal conductance characterization of a pressed copper rope strap between 0.13 K and 10 K}",
      journal = {Cryogenics},
         year = 2017,
        month = sep,
       volume = {86},
        pages = {17-21},
          doi = {10.1016/j.cryogenics.2017.07.001},
       adsurl = {https://ui.adsabs.harvard.edu/abs/2017Cryo...86...17D}
}

@article{Dhuley2019,
title = {Pressed copper and gold-plated copper contacts at low temperatures – A review of thermal contact resistance},
journal = {Cryogenics},
volume = {101},
pages = {111-124},
year = {2019},
issn = {0011-2275},
doi = {https://doi.org/10.1016/j.cryogenics.2019.06.008},
url = {https://www.sciencedirect.com/science/article/pii/S0011227519300943},
author = {R.C. Dhuley},
}

@ARTICLE{Didschuns2004,
       author = {{Didschuns}, I. and {Woodcraft}, A.~L. and {Bintley}, D. and {Hargrave}, P.~C.},
        title = "{Thermal conductance measurements of bolted copper to copper joints at sub-Kelvin temperatures}",
      journal = {Cryogenics},
         year = 2004,
        month = may,
       volume = {44},
       number = {5},
        pages = {293-299},
          doi = {10.1016/j.cryogenics.2003.11.010},
       adsurl = {https://ui.adsabs.harvard.edu/abs/2004Cryo...44..293D}
}

@INPROCEEDINGS{Doi2002,
       author = {{Doi}, Y. and {Yamamoto}, A. and {Makida}, Y. and {Kondo}, Y. and {Kawai}, M. and {Aoki}, K. and {Haruyama}, T. and {Kondo}, T. and {Mizumaki}, S. and {Wachi}, Y. and {Mine}, S. and {Haug}, F. and {Delruelle}, N. and {Passardi}, G. and {ten Kate}, H.},
        title = "{Performance of a proximity cryogenic system for the ATLAS central solenoid magnet}",
    booktitle = {Advances in Cryogenic Engineering CEC},
         year = 2002,
       series = {American Institute of Physics Conference Series},
       volume = {47},
        month = may,
    publisher = {AIP},
        pages = {47-54},
          doi = {10.1063/1.1472003},
       adsurl = {https://ui.adsabs.harvard.edu/abs/2002AIPC..613...47D}
}

@article{Dong2023,
title = {Contribution of grain boundary to strength and electrical conductivity of annealed copper wires},
journal = {Journal of Materials Research and Technology},
volume = {26},
pages = {1459-1468},
year = {2023},
issn = {2238-7854},
doi = {https://doi.org/10.1016/j.jmrt.2023.08.012},
url = {https://www.sciencedirect.com/science/article/pii/S2238785423018239},
author = {Liming Dong and Fei Yang and Tianbo Yu and Ning Zhang and Xuefeng Zhou and Zonghan Xie and Feng Fang}
}

@ARTICLE{Galitzki2024,
       author = {{Galitzki}, Nicholas and {Tsan}, Tran and {Spisak}, Jake and {Randall}, Michael and {Silva-Feaver}, Max and {Seibert}, Joseph and {Lashner}, Jacob and {Adachi}, Shunsuke and {Adkins}, Sean M. and {Alford}, Thomas and {Arnold}, Kam and {Ashton}, Peter C. and {Austermann}, Jason E. and {Baccigalupi}, Carlo and {Bazarko}, Andrew and {Beall}, James A. and {Bhimani}, Sanah and {Bixler}, Bryce and {Coppi}, Gabriele and {Corbett}, Lance and {Crowley}, Kevin D. and {Crowley}, Kevin T. and {Day-Weiss}, Samuel and {Devlin}, Mark J. and {Dicker}, Simon and {DiGia}, Brooke and {Dow}, Peter N. and {Duell}, Cody J. and {Duff}, Shannon M. and {Gerras}, Remington G. and {Groh}, John C. and {Gudmundsson}, Jon E. and {Harrington}, Kathleen and {Hasegawa}, Masaya and {Healy}, Erin and {Henderson}, Shawn W. and {Hubmayr}, Johannes and {Iuliano}, Jeffrey and {Johnson}, Bradley R. and {Keating}, Brian and {Keller}, Ben and {Kiuchi}, Kenji and {Kofman}, Anna M. and {Koopman}, Brian J. and {Kusaka}, Akito and {Lee}, Adrian T. and {Lew}, Richard A. and {Lin}, Lawrence T. and {Link}, Michael J. and {Lucas}, Tammy J. and {Lungu}, Marius and {Mangu}, Aashrita and {McMahon}, Jeffrey J. and {Miller}, Amber D. and {Moore}, Jenna E. and {Morshed}, Magdy and {Nakata}, Hironobu and {Nati}, Federico and {Newburgh}, Laura B. and {Nguyen}, David V. and {Niemack}, Michael D. and {Page}, Lyman A. and {Sakaguri}, Kana and {Sakurai}, Yuki and {Sathyanarayana Rao}, Mayuri and {Saunders}, Lauren J. and {Shroyer}, Jordan E. and {Sugiyama}, Junna and {Tajima}, Osamu and {Takeuchi}, Atsuto and {Bua}, Refilwe Tanah and {Teply}, Grant and {Terasaki}, Tomoki and {Ullom}, Joel N. and {Van Lanen}, Jeffrey L. and {Vavagiakis}, Eve M. and {Vissers}, Michael R. and {Walters}, Liam and {Wang}, Yuhan and {Xu}, Zhilei and {Yamada}, Kyohei and {Zheng}, Kaiwen},
        title = "{The Simons Observatory: Design, Integration, and Testing of the Small Aperture Telescopes}",
      journal = {The Astrophysical Journal Supplement},
         year = 2024,
        month = oct,
       volume = {274},
       number = {2},
          eid = {33},
        pages = {33},
          doi = {10.3847/1538-4365/ad64c9},
archivePrefix = {arXiv},
       eprint = {2405.05550},
 primaryClass = {astro-ph.IM},
       adsurl = {https://ui.adsabs.harvard.edu/abs/2024ApJS..274...33G}
}

@ARTICLE{Gmelin1999,
       author = {{Gmelin}, E. and {Asen-Palmer}, M. and {Reuther}, M. and {Villar}, R.},
        title = "{REVIEW ARTICLE: Thermal boundary resistance of mechanical contacts between solids at sub-ambient temperatures}",
      journal = {Journal of Physics D Applied Physics},
         year = 1999,
        month = mar,
       volume = {32},
       number = {6},
        pages = {R19-R43},
          doi = {10.1088/0022-3727/32/6/004},
       adsurl = {https://ui.adsabs.harvard.edu/abs/1999JPhD...32R..19G}
}

@INPROCEEDINGS{Homan2014,
       author = {{Homan}, J. and {Redman}, R. and {Ganni}, V. and {Sidi-Yekhlef}, A. and {Knudsen}, P. and {Norton}, R. and {Lauterbach}, J. and {Linza}, R. and {Vargas}, G.},
        title = "{Commissioning of a 20 K helium refrigeration system for NASA-JSC Chamber-A}",
    booktitle = {Advances in Cryogenic Engineering: Transactions of the Cryogenic Engineering Conference - CEC},
         year = 2014,
       series = {American Institute of Physics Conference Series},
       volume = {1573},
        month = jan,
    publisher = {AIP},
        pages = {307-314},
          doi = {10.1063/1.4860716},
       adsurl = {https://ui.adsabs.harvard.edu/abs/2014AIPC.1573..307H}
}

@book{Humphreys2024,
    author = {{Humphreys}, F.J. and {Hatherly}, M.},
    title = {Recrystallization and Related Annealing Phenomena},
    publisher = {Pergamon},
    year = {2024}
}

@book{hust_lankford_1984,
  author       = {Hust, J. G. and Lankford, A. B.},
  title        = {Thermal Conductivity of Aluminum, Copper, Iron, and Tungsten for Temperatures from 1 K to the Melting Point},
  address      = {Chemical Engineering Science Division, National Engineering Laboratory, National Bureau of Standards, U.S. Department of Commerce, Boulder, Colorado 80303},
  institution  = {National Bureau of Standards Publications},
  note         = {NBSIR 84-3007},
  year         = {1984},
  month        = jun
}

@INPROCEEDINGS{Karasev2012,
       author = {{Karasev}, Y.~V. and {Pantsyrny}, V.~I. and {Novosiliva}, D.~S. and {Polikarpova}, M.~V. and {Potanina}, L.~V. and {Gubkin}, I.~N. and {Malchenkov}, A.~V.},
        title = "{The effect of final processing steps on RRR of NbTi strands for ITER PF coils}",
    booktitle = {Advances in Cryogenic Engineering: Transactions of the International Cryogenic Materials Conference - ICMC, Volume 58},
         year = 2012,
       editor = {{Balachandran}, U. and {Amm}, Kathleen and {Ghosh}, Arup and {Hartwig}, K. Ted and {Lee}, Peter and {Osofsky}, Michael and {Pamidi}, Sastry and {Park}, Chan and {Stautner}, Wolfgang and {Sumption}, Michael and {Walsh}, Robert and {Zhang}, Tao},
       series = {American Institute of Physics Conference Series},
       volume = {58},
        month = jun,
    publisher = {AIP},
        pages = {182-189},
          doi = {10.1063/1.4712095},
       adsurl = {https://ui.adsabs.harvard.edu/abs/2012AIPC.1435..182K}
}

@INPROCEEDINGS{Koopman2020,
       author = {Brian J. Koopman and Jack Lashner and Lauren J. Saunders and Matthew Hasselfield and Tanay Bhandarkar and Sanah Bhimani and Steve K. Choi and Cody J. Duell and Nicholas Galitzki and Kathleen Harrington and Adam D. Hincks and Shuay-Pwu Patty Ho and Laura Newburgh and Christian L. Reichardt and Joseph Seibert and Jacob Spisak and Benjamin Westbrook and Zhilei Xu and Ningfeng Zhu},
        title = {{The Simons Observatory: overview of data acquisition, control, monitoring, and computer infrastructure}},
       volume = {11452},
    booktitle = {Software and Cyberinfrastructure for Astronomy VI},
       editor = {Juan C. Guzman and Jorge Ibsen},
 organization = {International Society for Optics and Photonics},
    publisher = {SPIE},
        pages = {1145208},
         year = {2020},
          doi = {10.1117/12.2561771},
          URL = {https://doi.org/10.1117/12.2561771}
}

@ARTICLE{McCarrick2021,
       author = {{McCarrick}, Heather and {Healy}, Erin and {Ahmed}, Zeeshan and {Arnold}, Kam and {Atkins}, Zachary and {Austermann}, Jason E. and {Bhandarkar}, Tanay and {Beall}, James A. and {Bruno}, Sarah Marie and {Choi}, Steve K. and {Connors}, Jake and {Cothard}, Nicholas F. and {Crowley}, Kevin D. and {Dicker}, Simon and {Dober}, Bradley and {Duell}, Cody J. and {Duff}, Shannon M. and {Dutcher}, Daniel and {Frisch}, Josef C. and {Galitzki}, Nicholas and {Gralla}, Megan B. and {Gudmundsson}, Jon E. and {Henderson}, Shawn W. and {Hilton}, Gene C. and {Ho}, Shuay-Pwu Patty and {Huber}, Zachary B. and {Hubmayr}, Johannes and {Iuliano}, Jeffrey and {Johnson}, Bradley R. and {Kofman}, Anna M. and {Kusaka}, Akito and {Lashner}, Jack and {Lee}, Adrian T. and {Li}, Yaqiong and {Link}, Michael J. and {Lucas}, Tammy J. and {Lungu}, Marius and {Mates}, J.~A.~B. and {McMahon}, Jeffrey J. and {Niemack}, Michael D. and {Orlowski-Scherer}, John and {Seibert}, Joseph and {Silva-Feaver}, Maximiliano and {Simon}, Sara M. and {Staggs}, Suzanne and {Suzuki}, Aritoki and {Terasaki}, Tomoki and {Thornton}, Robert and {Ullom}, Joel N. and {Vavagiakis}, Eve M. and {Vale}, Leila R. and {Van Lanen}, Jeff and {Vissers}, Michael R. and {Wang}, Yuhan and {Wollack}, Edward J. and {Xu}, Zhilei and {Young}, Edward and {Yu}, Cyndia and {Zheng}, Kaiwen and {Zhu}, Ningfeng},
        title = "{The Simons Observatory Microwave SQUID Multiplexing Detector Module Design}",
      journal = {The Astrophysical Journal},
         year = 2021,
        month = nov,
       volume = {922},
       number = {1},
          eid = {38},
        pages = {38},
          doi = {10.3847/1538-4357/ac2232},
archivePrefix = {arXiv},
       eprint = {2106.14797},
 primaryClass = {astro-ph.IM},
       adsurl = {https://ui.adsabs.harvard.edu/abs/2021ApJ...922...38M}
}

@ARTICLE{Nachman2026,
  author={Nachman, Henry E. and Desai, Oorie and Galitzki, Nicholas and Lee, Daniel and Lloyd, JB and Nandi, Tannishtha and Pagni, Ani and Radebaugh, Ray and Shaw, Elle C.},
  journal={IEEE Transactions on Applied Superconductivity}, 
  title={Cryogenic Materials Repository: A Public Resource and New Measurements for Cryogenic Research Applications}, 
  year={2026},
  volume={36},
  number={6},
  pages={9002607-9002607},
  doi={10.1109/TASC.2026.3713084}}

@article{OrlowskiScherer27,
	author = {{Orlowski-Scherer}, John},
	title = {The Simons Observatory: Commissioning of the Large Aperture Telescope},
	journal = {The Astrophysical Journal},
	year = "In Prep. 2027",
}

@book{PhysicalMetallurgy,
    author = {{Cahn}, Robert W. and {Haasen}, Peter},
    title = {Physical Metallurgy},
    publisher = {North Holland},
    year = 1996,
    edition = {Fourth},
}

@article{rauta2016,
    author = {{Rauta}, Veijo and {Cingi}, Celal and {Orkas}, Juhani},
    title = {Effect of Annealing and Metallurgical Treatments on Thermal Conductivity of Aluminium Alloys},
    journal = {Internaional Journal of Metalcasting},
    year = 2016,
    doi = {https://doi.org/10.1007/s40962-015-0017-z},
}

@book{Silberschmidt2024,
  author = {Vadim Silberschmidt},
  title = {Comprehensive Mechanics of Materials},
  year = {2024},
  publisher = {Elsevier},
  language = {en},
}

@techreport{simon1992,
  author       = {Simon, N J and Drexler, E S and Reed, R P},
  title        = {Properties of copper and copper alloys at cryogenic temperatures. Final report},
  institution  = {National Inst. of Standards and Technology (MSEL), Boulder, CO (United States). Materials Reliability Div.},
  doi          = {10.2172/5340308},
  url          = {https://www.osti.gov/biblio/5340308},
  place        = {United States},
  year         = {1992},
  month        = {02}}

@ARTICLE{SOgoals,
       author = {{Ade}, Peter and {Aguirre}, James and {Ahmed}, Zeeshan and {Aiola}, Simone and {Ali}, Aamir and {Alonso}, David and {Alvarez}, Marcelo A. and {Arnold}, Kam and {Ashton}, Peter and {Austermann}, Jason and {Awan}, Humna and {Baccigalupi}, Carlo and {Baildon}, Taylor and {Barron}, Darcy and {Battaglia}, Nick and {Battye}, Richard and {Baxter}, Eric and {Bazarko}, Andrew and {Beall}, James A. and {Bean}, Rachel and {Beck}, Dominic and {Beckman}, Shawn and {Beringue}, Benjamin and {Bianchini}, Federico and {Boada}, Steven and {Boettger}, David and {Bond}, J. Richard and {Borrill}, Julian and {Brown}, Michael L. and {Bruno}, Sarah Marie and {Bryan}, Sean and {Calabrese}, Erminia and {Calafut}, Victoria and {Calisse}, Paolo and {Carron}, Julien and {Challinor}, Anthony and {Chesmore}, Grace and {Chinone}, Yuji and {Chluba}, Jens and {Cho}, Hsiao-Mei Sherry and {Choi}, Steve and {Coppi}, Gabriele and {Cothard}, Nicholas F. and {Coughlin}, Kevin and {Crichton}, Devin and {Crowley}, Kevin D. and {Crowley}, Kevin T. and {Cukierman}, Ari and {D'Ewart}, John M. and {D{\"u}nner}, Rolando and {de Haan}, Tijmen and {Devlin}, Mark and {Dicker}, Simon and {Didier}, Joy and {Dobbs}, Matt and {Dober}, Bradley and {Duell}, Cody J. and {Duff}, Shannon and {Duivenvoorden}, Adri and {Dunkley}, Jo and {Dusatko}, John and {Errard}, Josquin and {Fabbian}, Giulio and {Feeney}, Stephen and {Ferraro}, Simone and {Flux{\`a}}, Pedro and {Freese}, Katherine and {Frisch}, Josef C. and {Frolov}, Andrei and {Fuller}, George and {Fuzia}, Brittany and {Galitzki}, Nicholas and {Gallardo}, Patricio A. and {Tomas Galvez Ghersi}, Jose and {Gao}, Jiansong and {Gawiser}, Eric and {Gerbino}, Martina and {Gluscevic}, Vera and {Goeckner-Wald}, Neil and {Golec}, Joseph and {Gordon}, Sam and {Gralla}, Megan and {Green}, Daniel and {Grigorian}, Arpi and {Groh}, John and {Groppi}, Chris and {Guan}, Yilun and {Gudmundsson}, Jon E. and {Han}, Dongwon and {Hargrave}, Peter and {Hasegawa}, Masaya and {Hasselfield}, Matthew and {Hattori}, Makoto and {Haynes}, Victor and {Hazumi}, Masashi and {He}, Yizhou and {Healy}, Erin and {Henderson}, Shawn W. and {Hervias-Caimapo}, Carlos and {Hill}, Charles A. and {Hill}, J. Colin and {Hilton}, Gene and {Hilton}, Matt and {Hincks}, Adam D. and {Hinshaw}, Gary and {Hlo{\v{z}}ek}, Ren{\'e}e and {Ho}, Shirley and {Ho}, Shuay-Pwu Patty and {Howe}, Logan and {Huang}, Zhiqi and {Hubmayr}, Johannes and {Huffenberger}, Kevin and {Hughes}, John P. and {Ijjas}, Anna and {Ikape}, Margaret and {Irwin}, Kent and {Jaffe}, Andrew H. and {Jain}, Bhuvnesh and {Jeong}, Oliver and {Kaneko}, Daisuke and {Karpel}, Ethan D. and {Katayama}, Nobuhiko and {Keating}, Brian and {Kernasovskiy}, Sarah S. and {Keskitalo}, Reijo and {Kisner}, Theodore and {Kiuchi}, Kenji and {Klein}, Jeff and {Knowles}, Kenda and {Koopman}, Brian and {Kosowsky}, Arthur and {Krachmalnicoff}, Nicoletta and {Kuenstner}, Stephen E. and {Kuo}, Chao-Lin and {Kusaka}, Akito and {Lashner}, Jacob and {Lee}, Adrian and {Lee}, Eunseong and {Leon}, David and {Leung}, Jason S. -Y. and {Lewis}, Antony and {Li}, Yaqiong and {Li}, Zack and {Limon}, Michele and {Linder}, Eric and {Lopez-Caraballo}, Carlos and {Louis}, Thibaut and {Lowry}, Lindsay and {Lungu}, Marius and {Madhavacheril}, Mathew and {Mak}, Daisy and {Maldonado}, Felipe and {Mani}, Hamdi and {Mates}, Ben and {Matsuda}, Frederick and {Maurin}, Lo{\"\i}c and {Mauskopf}, Phil and {May}, Andrew and {McCallum}, Nialh and {McKenney}, Chris and {McMahon}, Jeff and {Meerburg}, P. Daniel and {Meyers}, Joel and {Miller}, Amber and {Mirmelstein}, Mark and {Moodley}, Kavilan and {Munchmeyer}, Moritz and {Munson}, Charles and {Naess}, Sigurd and {Nati}, Federico and {Navaroli}, Martin and {Newburgh}, Laura and {Nguyen}, Ho Nam and {Niemack}, Michael and {Nishino}, Haruki and {Orlowski-Scherer}, John and {Page}, Lyman and {Partridge}, Bruce and {Peloton}, Julien and {Perrotta}, Francesca and {Piccirillo}, Lucio and {Pisano}, Giampaolo and {Poletti}, Davide and {Puddu}, Roberto and {Puglisi}, Giuseppe and {Raum}, Chris and {Reichardt}, Christian L. and {Remazeilles}, Mathieu and {Rephaeli}, Yoel and {Riechers}, Dominik and {Rojas}, Felipe and {Roy}, Anirban and {Sadeh}, Sharon and {Sakurai}, Yuki and {Salatino}, Maria and {Sathyanarayana Rao}, Mayuri and {Schaan}, Emmanuel and {Schmittfull}, Marcel and {Sehgal}, Neelima and {Seibert}, Joseph},
        title = "{The Simons Observatory: science goals and forecasts}",
      journal = {Journal of Cosmology and Astroparticle Physics},
         year = 2019,
        month = feb,
       volume = {2019},
       number = {2},
          eid = {056},
        pages = {056},
          doi = {10.1088/1475-7516/2019/02/056},
archivePrefix = {arXiv},
       eprint = {1808.07445},
 primaryClass = {astro-ph.CO},
       adsurl = {https://ui.adsabs.harvard.edu/abs/2019JCAP...02..056A}
}

@ARTICLE{Stevens2020,
       author = {{Stevens}, Jason R. and {Cothard}, Nicholas F. and {Vavagiakis}, Eve M. and {Ali}, Aamir and {Arnold}, Kam and {Austermann}, Jason E. and {Choi}, Steve K. and {Dober}, Bradley J. and {Duell}, Cody and {Duff}, Shannon M. and {Hilton}, Gene C. and {Ho}, Shuay-Pwu Patty and {Hoang}, Thuong D. and {Hubmayr}, Johannes and {Lee}, Adrian T. and {Mangu}, Aashrita and {Nati}, Federico and {Niemack}, Michael D. and {Raum}, Christopher and {Renzullo}, Mario and {Salatino}, Maria and {Sasse}, Trevor and {Simon}, Sara M. and {Staggs}, Suzanne and {Suzuki}, Aritoki and {Truitt}, Patrick and {Ullom}, Joel and {Vivalda}, John and {Vissers}, Michael R. and {Walker}, Samantha and {Westbrook}, Benjamin and {Wollack}, Edward J. and {Xu}, Zhilei and {Yohannes}, Daniel},
        title = "{Characterization of Transition Edge Sensors for the Simons Observatory}",
      journal = {Journal of Low Temperature Physics},
         year = 2020,
        month = feb,
       volume = {199},
       number = {3-4},
        pages = {672-680},
          doi = {10.1007/s10909-020-02375-9},
archivePrefix = {arXiv},
       eprint = {1912.00860},
 primaryClass = {astro-ph.IM},
       adsurl = {https://ui.adsabs.harvard.edu/abs/2020JLTP..199..672S}
}

@article{Woodcraft2005,
title = {Recommended values for the thermal conductivity of aluminium of different purities in the cryogenic to room temperature range, and a comparison with copper},
journal = {Cryogenics},
volume = {45},
number = {9},
pages = {626-636},
year = {2005},
issn = {0011-2275},
doi = {https://doi.org/10.1016/j.cryogenics.2005.06.008},
url = {https://www.sciencedirect.com/science/article/pii/S0011227505000895},
author = {Adam L. Woodcraft}
}

@INPROCEEDINGS{Woodcraft2009,
       author = {{Woodcraft}, Adam L. and {Gray}, Adam},
        title = "{A low temperature thermal conductivity database}",
    booktitle = {The Thirteenth International Workshop on Low Temperature Detectors - LTD13},
         year = 2009,
       editor = {{Young}, Betty and {Cabrera}, Blas and {Miller}, Aaron},
       series = {American Institute of Physics Conference Series},
       volume = {1185},
        month = dec,
    publisher = {AIP},
        pages = {681-684},
          doi = {10.1063/1.3292433},
       adsurl = {https://ui.adsabs.harvard.edu/abs/2009AIPC.1185..681W}
}

@ARTICLE{Zhu2021,
       author = {{Zhu}, Ningfeng and {Bhandarkar}, Tanay and {Coppi}, Gabriele and {Kofman}, Anna M. and {Orlowski-Scherer}, John L. and {Xu}, Zhilei and {Adachi}, Shunsuke and {Ade}, Peter and {Aiola}, Simone and {Austermann}, Jason and {Bazarko}, Andrew O. and {Beall}, James A. and {Bhimani}, Sanah and {Bond}, J. Richard and {Chesmore}, Grace E. and {Choi}, Steve K. and {Connors}, Jake and {Cothard}, Nicholas F. and {Devlin}, Mark and {Dicker}, Simon and {Dober}, Bradley and {Duell}, Cody J. and {Duff}, Shannon M. and {D{\"u}nner}, Rolando and {Fabbian}, Giulio and {Galitzki}, Nicholas and {Gallardo}, Patricio A. and {Golec}, Joseph E. and {Haridas}, Saianeesh K. and {Harrington}, Kathleen and {Healy}, Erin and {Ho}, Shuay-Pwu Patty and {Huber}, Zachary B. and {Hubmayr}, Johannes and {Iuliano}, Jeffrey and {Johnson}, Bradley R. and {Keating}, Brian and {Kiuchi}, Kenji and {Koopman}, Brian J. and {Lashner}, Jack and {Lee}, Adrian T. and {Li}, Yaqiong and {Limon}, Michele and {Link}, Michael and {Lucas}, Tammy J. and {McCarrick}, Heather and {Moore}, Jenna and {Nati}, Federico and {Newburgh}, Laura B. and {Niemack}, Michael D. and {Pierpaoli}, Elena and {Randall}, Michael J. and {Sarmiento}, Karen Perez and {Saunders}, Lauren J. and {Seibert}, Joseph and {Sierra}, Carlos and {Sonka}, Rita and {Spisak}, Jacob and {Sutariya}, Shreya and {Tajima}, Osamu and {Teply}, Grant P. and {Thornton}, Robert J. and {Tsan}, Tran and {Tucker}, Carole and {Ullom}, Joel and {Vavagiakis}, Eve M. and {Vissers}, Michael R. and {Walker}, Samantha and {Westbrook}, Benjamin and {Wollack}, Edward J. and {Zannoni}, Mario},
        title = "{The Simons Observatory Large Aperture Telescope Receiver}",
      journal = {The Astrophysical Journal Supplement},
         year = 2021,
        month = sep,
       volume = {256},
       number = {1},
          eid = {23},
        pages = {23},
          doi = {10.3847/1538-4365/ac0db7},
archivePrefix = {arXiv},
       eprint = {2103.02747},
 primaryClass = {astro-ph.IM},
       adsurl = {https://ui.adsabs.harvard.edu/abs/2021ApJS..256...23Z}
}

\end{document}